\documentclass[10pt,letterpaper]{article}
\usepackage{opex3}

\begin{document}

\title{Enhancing  and controlling single-atom high-harmonic generation
  spectra: a time-dependent density-functional scheme}

\author{Alberto Castro$^{1,2,*}$, Angel Rubio$^3$, and E. K. U. Gross$^4$}

\address{
$^1$ARAID Foundation, Edificio CEEI, Mar{\'{\i}}a de Luna s/n, 50018 Zaragoza Spain\\
$^2$Institute for Biocomputation and Physics of Complex
Systems (BIFI), and Zaragoza Center for Advanced Modelling (ZCAM), University
of Zaragoza, 50018 Zaragoza, Spain\\
$^3$Nano-Bio Spectroscopy Group and ETSF Scientific Development Centre, Departamento de F\'isica de Materiales, Centro de F\'isica de Materiales CSIC-UPV/EHU-MPC and DIPC, Universidad del Pa\'is Vasco UPV/EHU, E-20018 San Sebasti\'an, Spain\\
$^4$Max-Planck Institut f{\"{u}}r Mikrostrukturphysik, Weinberg 2, D-06120
Halle, Germany
}

\email{$^*$acastro@bifi.es}



\begin{abstract}
  High harmonic generation (HHG) provides a flexible framework for the
  development of coherent light sources in the extreme-ultraviolet and
  soft x-ray regimes.  However it suffers from low conversion
  efficiencies as the control of the HHG spectral and temporal
  characteristics requires manipulating electron trajectories on
  attosecond time scale. The phase matching mechanism has been
  employed to selectively enhance specific quantum paths leading to
  HHG. A few important fundamental questions remain open, among those
  how much of the enhancement can be achieved by the single-emitter
  and what is the role of correlations (or the electronic structure)
  in the selectivity and control of HHG generation. Here we address
  those questions by examining computationally the possibility of
  optimizing the HHG spectrum of isolated Hydrogen and Helium atoms by
  shaping the slowly varying envelope of a 800 nm, 200-cycles long
  laser pulse.  The spectra are computed with a fully quantum
  mechanical description, by explicitly computing the time-dependent
  dipole moment of the systems using a first-principles time-dependent
  density-functional approach (exact for the case of H). The sought
  optimization corresponds to the selective enhancement of single
  harmonics, which we find to be significant. This selectivity is
  entirely due to the single atom response, and not due to any
  propagation or phase-matching effect. In fact, this single-emitter
  enhancement adds to the phase-matching techniques to achieving even
  larger HHG enhancement factors. Moreover, we see that the electronic
  correlation plays a role in the determining the degree of
  optimization that can be obtained.
\end{abstract}




\section{Introduction}

At sufficiently high intensities, matter no longer reacts linearly to light,
and may re-emit at integer multiples (\emph{harmonics}) of the frequency of
the incoming source~\cite{franken1961}. According to perturbation theory, the
intensity of the harmonics decreases exponentially with their order. However,
the spectrum of atoms and molecules exposed to very intense, typically
infrared, laser pulses was found to present unexpectedly high
harmonics~\cite{mcpherson1987,ferray1988}, and its shape was observed to have
a plateau extending non-perturbatively over many orders of magnitude -- a
process known as \emph{high harmonic generation}
(HHG)~\cite{Smirnova2013,Ishikawa2010}.  The light emitted in this manner is
coherent and may reach the extreme ultraviolet and soft X-ray frequency regime
opening the path towards the coherent manipulation and control of matter at
its natural time scale. These properties can be of paramount importance for
many technological and scientific purposes in ultrafast science -- most
notably, the generation of attosecond pulse trains or single isolated
attosecond pulses, or the external seeding of free-electron
lasers~\cite{hentschel2001,wirth2011,Brabec2000}. These advances allow to
follow the electron dynamics~\cite{drescher2002}.  Unsurprisingly, a big
effort has been devoted to first understanding the underlying physics, and
then to controlling and fine-tuning the efficiency and spectral
characteristics of the harmonic radiation. The latter
can be done by modifying the non-linear medium, or by post-processing the
signal with filters, gratings, etc. However, one advantageous alternative is
to modify the characteristics of the parent pulse, which obviously will modify
the spectral outcome.  The most obvious manner of doing this is by
systematically varying the defining parameters of this parent
pulse~\cite{Lee2001,Chang1997,Kim2003}. However, the current availability of
advanced pulse shaping tools~\cite{weiner2000}, together with the development
of closed-loop quantum control techniques~\cite{Brif2010}, provides a superior
optimization alternative~\cite{winterfeldt2008}.  In this manner, the
successful selective enhancement of harmonic orders could be achieved when
using a hollow fiber container for the generating
medium~\cite{bartels2000,bartels2001,pfeifer2005}. Gas jet (\emph{free
  focusing}) geometries were also
employed~\cite{reitze2004,villoresi2004,walter2006}, but although some degree
of control could be achieved (for example, the extension of the cut-off
frequency), the very selective order enhancement or depletion obtained with
the hollow fibers was not observed. This fact seems to imply that this type of
selective enhancement cannot be explained from the single-atom response only;
instead, the propagation effects present in the capillary set-up apparently
play a fundamental role. In this context. quasiphase matching (QPM) approach
is commonly used to achieve independent phase control between multiple
high-harmonics~\cite{Paul2003}.

However, a full interpretation of the optimisation mechanisms can only be
achieved with theoretical input, for which purpose one may utilise quantum
simulations in combination with the theoretical branch of quantum optimal
control~\cite{doi:10.1088/0953-4075/40/18/R01,Brif2010} (QOCT).  Recently,
Schaefer and Kosloff~\cite{Schaefer2012, schaefer} have addressed this task,
showing the possibility of enhancing the emission at desired frequencies for
simple few level systems and one-dimensional one.electron system.  Here we
address, by first principles simulations based on time-dependent density
functional theory, the role of many electron interactions in the high harmonic
generation, and provide compelling evidence that a single-atom HHG emission can
be enhanced by few orders of magnitude in a controlled manner, with standard
laser shaping techniques available in many experimental labs.

The \emph{three-step} model successfully describes the
key features of HHG~\cite{krause1992,schafer-1993,corkum1994}, at least
qualitatively. It combines a quantum description of the ionisation and
recombination of the electrons, with a classical description of the
intermediate electronic propagation.  Lewenstein \emph{et
  al}~\cite{lewenstein1994} developed an approximate, mostly analytical,
quantum description based on the \emph{strong field approximation} (SFA): it
neglects the contribution of excited bound states, the depletion of the ground
state, and considers the continuum electrons to be free of the influence of
the parent ion.  A more precise approach consists of propagating
Schr{\"{o}}dinger's equation~\cite{Kulander1993,Tong1997,bandrauk2009}, an
expensive method that quickly becomes prohibitive as we increase the number of
electrons. For one-electron problems the approach is perfectly feasible, and
this fact has encouraged the use of the \emph{single active electron}
approximation (SAE), which assumes that only one electron is significantly
disturbed by the field, and its evolution may be computed on the combination
of the laser field and the potential originating by the parent ion.

This single electron picture is commonly used to describe recollision
processes and HHG in atoms and relies on the fact that under HHG conditions
there is one electron being ionised. However this doesn't imply the other
electrons do not play a role. There is indeed no formal justification for the
use of the SAE and in fact, many-body effects have been shown recently to play
an important role in HHG providing an explanation of why heavier atoms emit
stronger HHG than lighter ones~\cite{Gordon2006} and the giant enhancement of
He HHG at 100~ev~\cite{Pabst2013}. However, the SAE has been successful in
explaining a few features of the HHG spectra such as the spectral cutoff, the
phase structure of the spectrum and the prediction of the generation of
attosecond pulses.

In spite of all those experimental and theoretical efforts, it is clear that
the topic of selective HHG generation deserves further microscopical analysis,
and in this work, we explore the optimisation possibilities of one and two
electron systems (the Hydrogen and the Helium atoms), isolating the single
atom response, so that we can learn how much selectivity in the HHG spectrum
can be obtained from isolated atoms that can nicely complement QPM schemes in
enhancing further the HHG selective emission. For this purpose, we employ a
global optimisation scheme that acts on the envelope of the generating pulse,
maintaining the fundamental frequency and minimising undesired ionisation (and
for molecules also dissociation) processes.  For the case of Helium, we report
results obtained both with the single active electron approximation, and with
time-dependent density-functional theory (TDDFT)~\cite{runge1984,tddft}, in
order to assess the influence of the electron-electron interaction in the
optimisation process.  As many-electron effects may be relevant, TDDFT appears
as the ideal framework to capture them in the HHG spectra (see for example
Ref.~\cite{ullrich1996}) as it combines a very good compromise between
accuracy and computational efficiency. The present optimisation scheme has
been implemented in the first-principles code {\tt
  octopus}~\cite{marques2003,castro2006}, that allows the treatment of more
complex molecular and extended systems.  However for the purposes of the
present work, it is better to stay at the simplest level of one and two
electron systems. Larger electronic systems would offer a wider range of
possibilities for HHG enhancement.

\section{Theory}

Within the dipole approximation and in the length gauge,
the experimentally measured harmonic spectrum can be theoretically
approximated by the following formula:
\begin{equation}
\label{eq:hhgfunction}
H(\omega) = \vert \int_0^T\!\!{\rm d}t\; \frac{\rm d^2}{{\rm d}t^2} \langle
\hat{\vec{\mu}}\rangle(t) e^{-{\rm i}\omega t}\vert^2\,,
\end{equation}
i.e. the power spectrum of the second derivative of the expectaction value of
the dipole moment $\hat{\vec{\mu}} = - \sum_{i=1}^N \hat{\vec{r}}_i$
(although see Ref.~\cite{Sundaram1990} for a discussion on the
pertinence of using, alternatively, the first derivative or the dipole moment
itself). This object is given by:
\begin{equation}
\label{eq:d2dmudt2}
\frac{\rm d^2}{{\rm d}t^2} \langle \hat{\vec{\mu}}\rangle(t) = 
\langle \sum_{i=1}^N\nabla v(\hat{\vec{r}}_i) \rangle + N\varepsilon(t)\vec{\pi}\,,
\end{equation}
where $v$ is the (static) ionic potential, $N$ is the number of electrons,
$\varepsilon(t)$ is the laser pulse electric field, and $\vec{\pi}$ is the
polarization vector (see the Methods appendix below for some extra details
about the theory).  Note that this
expression can be read as both the acceleration of the electronic system, and
as the corresponding back-reaction of the nucleus (or nuclear center of mass,
if we are dealing with a molecule). This is not surprising since the
electromagnetic emission must be related with a charge acceleration. The
expression corresponds, except for the mass factor, with the classical
\emph{force} acting on the nucleus, considered as a point particle. We will
therefore rewrite Eq.~(\ref{eq:hhgfunction}) as:
\begin{equation}
H(\omega) = \vert \vec{f}(\omega)\vert^2\,.
\end{equation}
where $\vec{f}(\omega) = \int_0^T\!\! \vec{f}(t)e^{-i\omega t})$ is the Fourier transform of:
\begin{equation}
\label{eq:ft}
\vec{f}(t) = 
\langle \sum_{i=1}^N\nabla v(\hat{\vec{r}}_i) \rangle +
N\varepsilon(t)\vec{\pi}\,.
\end{equation}

From a TDDFT perspective, the use of this force functional is convenient since
it can be explicitly written as a density functional:
\begin{equation}
\label{eq:d2dmu}
\vec{f}(t) = 
\int\!\!{\rm
  d}^3r\; n(\vec{r},t) \nabla v(\vec{r})
+ N\varepsilon(t)\vec{\pi}\,.
\end{equation}

Usually, the electric field $\varepsilon(t)$ is factorised into a sinusoidal
function determining the fundamental frequency $\omega_0$, and an
\emph{envelope} function $f$ that determines the overall laser-pulse shape:
\begin{equation}
\varepsilon(t) = f(t)\sin(\omega_0 t)\,.
\end{equation}
This factorisation -- and the concept of a \emph{fundamental frequency} -- is
meaningful for long and quasi-monochromatic pulses, but as the technology has
reached the optical period limit, it has started to lose its
relevance. Nevertheless, the existence of a fundamental frequency is implicit
when speaking of harmonics, which are defined as radiation at integer
multiples of precisely that frequency. These will only be well defined if
the envelope function is \emph{smooth} compared to the sinusoidal term, i.e.
its frequencies are much lower than $\omega_0$.

Therefore, in this work, we investigate the possibility of manipulating the
envelope function $f$, leaving the sinusoidal factor $\sin(\omega_0 t)$
unchanged, in order to influence the shape of the HHG spectrum. This
manipulation cannot be unconstrained, as the envelope must be composed of
frequencies much lower than $\omega_0$. Moreover, we have searched for
solutions that preserve the fluence or total integrated energy of the pulse:
\begin{equation}
\label{eq:fluence}
\overline{I} = \int\!\!{\rm d}t\; \varepsilon^2(t)\,.
\end{equation}

This type of requirement of a specific structure for the solution field (in
terms of frequencies, fluence, etc.) can be respected following essentially
two routes: by imposing penalties on undesired features of the pulses in the
definition of the optimising function, or by constraining from the beginning
the search space. This latter option can be achieved by establishing a
parametrisation of the control field (in this case, the envelope) that
enforces the required condition, and is the route that we have chosen for this
work. The search for the optimum is in this manner performed in the space of
parameters that determine the control field; the remaining necessary
ingredient is the definition of a \emph{merit} function that encodes the
physical requirements.  Moreover, the assumption of low frequencies for $f$ implies that
the spectrum of $\varepsilon$ is concentrated around $\omega_0$. Therefore, the
$N\varepsilon(t)\vec{\pi}$ term in Eqs.~(\ref{eq:d2dmudt2}), (\ref{eq:ft}), and
(\ref{eq:d2dmu}) does not contribute to the HHG spectrum in the region we are
interested in and in the following we will safely ignore it. 

We have shown how the HHG spectrum may be explicitly computed solely in terms
of the system electronic density $n(\vec{r},t)$. For systems with more than
one electron, this fact is convenient since it allows to use time-dependent
density-functional theory~\cite{runge1984, tddft} (TDDFT) (see Methods). One
can substitute the propagation of the real interacting system by the
propagation of a system of fictitious non-interacting electrons whose density
is however identical to that of the real one, despite the fact that its wave
function is a single Slater determinant.

Hereafter, we will restrict the discussion to one and two-electron systems,
the extension to systems with larger number of electrons is straightforward in
the TDDFT framework.  The one-electron case obviously does not need a TDDFT
treatment, although it may be treated as such by considering one single
occupied orbital.  For such one-orbital problem, the exchange and correlation
potential must cancel the Hartree term:
\begin{equation}
v_{\rm xc}[n](\vec{r},t) = - v_{\rm H}[n](\vec{r},t)\,,
\end{equation}
so that the resulting equation reduces to the initial
Schr{\"{o}}dinger equation. For two-electron systems, we use the
exact-exchange approximation (EXX) to the xc term, which for this two-electron
case amounts to setting:
\begin{equation}
v_{\rm xc}[n](\vec{r},t) = - \frac{1}{2}v_{\rm H}[n](\vec{r},t)\,,
\end{equation}
Note that in this form TDDFT is identical to time-dependent Hartree-Fock that
provides a food description of the non-linear properties of two-electron
systems except for the description of charge-transfer excitations (see for
example Ref.~\cite{Fuks2013}).

The electric field amplitude will be determined by the specification of
a set of $M$ parameters $u_1,\dots,u_M \equiv u: \varepsilon(t) = \varepsilon[u](t)$.
The evolution of the TDKS system is in consequence also governed by the choice of
parameters $u$, i.e. the orbitals and density are functionals of the
parameters: $u \rightarrow \varphi[u]$, $u \rightarrow n[u].$ We may then use
the tools of QOCT to find the set $u$ that maximizes a given target function
$G$, defined in terms of a functional of the density of the system, i.e.:
\begin{equation}
G[u] = \tilde{F}[n[u]]\,.
\end{equation}
This functional $\tilde{F}$ is designed to favour the desired behaviour of the
system (in this case, a certain form of the HHG spectrum, to be detailed
below). Note that it is defined in terms of the system density, and not in
terms of the full many-body wave function. This definition ensures that the substitution
of the real by the Kohn-Sham system in the optimization entails no further
approximation. The theory must however be developed in terms of a functional
of the Kohn-Sham orbitals, which can be easily defined as:
\begin{equation}
F[\varphi] = \tilde{F}[\mu\varphi^*\varphi]\,,
\end{equation}
where $\mu$ is the occupation of the orbital, i.e. one or two for one- or
two-electron calculations, respectively.

We must now choose a form for $F$ in such a way that its maximization leads to
the desired HHG optimization, namely the selective increase of one harmonic
peak -- that should leave the neighboring ones as low as possible. There is
substantial liberty to design $F$, and it is not evident what functional form
should lead to better results.
\begin{equation}
F[\varphi]  = \sum_{k} \alpha_k  H[\varphi](k\omega_0) \,,
\end{equation}
where $\alpha_k$ takes a positive value for the harmonic to be enhanced, and
negative values for the ones that we wish to reduce. However, this choice
proved to be problematic, since the modulation of the source signal with the
envelope function leads to displacements, sometimes substantial, of the
harmonic peaks with respect to the precise integer multiples $k\omega_0$. A
general definition that solves this problem (and that includes the previous one as
a particular case), is:
\begin{equation}
\label{eq:hhgtarget1}
F[\varphi] =  \int\!\!{\rm d}\omega \alpha(\omega)H[\varphi](\omega) 
=
\int\!\!{\rm d}\omega \alpha(\omega)
\vert \vec{f}[\varphi](\omega)\vert^2\,,
\end{equation}
where we have made explicit the fact that both $H$ and $\vec{f}$, defined in
Eqs.~(\ref{eq:hhgfunction}) and (\ref{eq:ft}) are functionals of the
time-dependent evolution for the system. The function $\alpha$ permits to
establish some finite window around each harmonic peak $k\omega_0$, that will
be positive for the harmonic orders that we want to enhance, and negative for
the ones that we want to reduce. Finally, a third option is to seek for the
maximum of the spectrum in these frequency windows around the harmonic orders,
i.e.:
\begin{equation}
\label{eq:hhgtarget2}
F[\varphi]  = \sum_{k} \alpha_k \max_{\omega \in [k\omega_0 - \beta, k\omega_0
    + \beta]} \lbrace \log_{10} H[\varphi](\omega)\rbrace \,,
\end{equation}
where the real number $\beta$ determines the size of the window.

Once the function $G$ has been defined (through the definition of
the target functional $F$), it remains to use some optimization algorithm
 to find the optimal $u$ set. There are numerous options, and we may
divide them on two groups, depending on whether or not they require the
computation of the gradient of $G$ -- in addition of the computation of the
function itself. The methods that employ the gradient are of course more
efficient, as long as this gradient can itself be computed efficiently. 
The simplest scheme is steepest descents, but one can also use conjugate
gradients or, in our case, the Broyden-Fletcher-Goldfarb-Shanno (GFBS)
quasi-Newton method.

For the function $G$, the gradient is given by~\cite{Castro2012a}:
\begin{equation}
\nabla G[u] = 
2 \int_0^T\!\!{\rm d}t\; \nabla \varepsilon[u](t)  {\rm Im} 
\langle \chi[u](t)\vert \hat{\vec{r}}\cdot\hat{\pi} \vert \varphi[u](t)\rangle\,.
\end{equation}
This expression uses an auxiliary orbital $\chi[u]$
defined by the following equations of motion:
\begin{eqnarray}
\nonumber
{\rm i}\frac{\partial}{\partial t}\chi[u](\vec{r},t) & = & 
-\frac{1}{2}\nabla^2 \chi[u](\vec{r},t)
+ v^*_{\rm KS}[n[u]](\vec{r},t)\chi[u](\vec{r},t)
\\\nonumber
 & & + \hat{K}[\varphi[u](t)] \chi[u](\vec{r},t)
\\
\label{eq:chit}
& & -i \frac{\delta F[\varphi[u]]}{\delta \varphi^*[u](\vec{r},t)}\,,
\\
\label{eq:bc-final}
\chi[u](\vec{r},T) & = & 0\,.
\end{eqnarray}
The operator $\hat{K}[\varphi[u][t]]$ is defined as:
\begin{equation}
\nonumber
\hat{K}[\varphi[u](t)] \chi[u](\vec{r},t)  =   
-4 {\rm i}\varphi[u](\vec{r},t) 
{\rm Im}
\int{\rm d}^3r'\;\chi^*[u](\vec{r}',t) f_{\rm Hxc}[n[u](t)](\vec{r},\vec{r}')
\varphi[u](\vec{r}',t)\,,
\end{equation}
where $f_{\rm Hxc}$ is the so-called \emph{kernel} of the Kohn-Sham
Hamiltonian, which, for our two-electron case treated within the EXX
approximation, is given by: $f_{\rm Hxc}[n](\vec{r},\vec{r}') = \frac{1}{2}\frac{1}{\vert \vec{r}-\vec{r}'\vert}$,
and is null for the one-electron case (zeroing the full $\hat{K}$ operator).

The functional derivative of $F$, needed in Eq.~(\ref{eq:chit}), for
the HHG target defined in Eq.~(\ref{eq:hhgtarget1}), is:
\begin{equation}
  \frac{\delta F}{\delta \varphi^*(\vec{r},t)} = 
\vec{g}[\varphi](t)\cdot\nabla v(\vec{r})\;\varphi(\vec{r},t)\,,
\end{equation}
where $\vec{g}[\varphi](t) = 2\mu \int\!\!{\rm d}\omega\;\alpha(\omega) {\rm
  Re}\left[ \vec{f}[\varphi](\omega)e^{-i\omega t}\right]$.

However, we cannot compute this functional derivative for the target defined
in Eq.~(\ref{eq:hhgtarget2}) due to the presence of the ``max'' function, at
least in a simple and efficient manner. In consequence, when using this target
definition we could not make use of any of the optimization algorithms that
make use of the gradient, and turned to the gradient-free NEWUOA
algorithm~\cite{Powell2008a}, which is a very efficient scheme for
optimization problems with a moderate number of degrees of freedom, such as
the ones treated here. 

In fact, for the optimizations attempted in this work, we observed numerically that the
target of Eq.~(\ref{eq:hhgtarget2}) provided much better results than the
target of Eq.~(\ref{eq:hhgtarget1}), and therefore we will only show below
gradient-free optimizations; in a forthcoming publication, where the target is
the HHG cut-off extension, we will present gradient-based optimizations based
on a target of the type given in Eq.~(\ref{eq:hhgtarget1}).

Therefore, it remains to specify the set of parameters $u$ that determine the envelope of
the electric fields. The requirements are: (i) the envelope should have a
given cut-off frequency; (ii) the field should smoothly approach zero at the
end points of the propagation time interval; (iii) the total integral of the
field should be zero, and (iv) the \emph{fluence} or total integrated
intensity of the pulse should have a constant pre-defined value. This last
condition is merely a choice, and not a physical constraint that
experimentalists face.

The first step to parametrize the applied time-dependent electroc field
 $\varepsilon(t)$ in order to enforce all these
constraints is to expand the envelope in a Fourier series:
\begin{equation}
f(t) = \sum_{i=1}^{2L} f_i g_i(t)\,,
\end{equation}
where
\begin{equation}
g_i(t) = \left\{\begin{array}{ll}
\sqrt{\frac{2}{T}} \cos\left( \frac{2\pi}{T}i t \right)\,&  (i = 1,\dots,L)
\\
\sqrt{\frac{2}{T}} \sin\left( \frac{2\pi}{T}(i-L) t \right)\,& (i = L+1,\dots,2L)
\end{array}\right.
\end{equation}
This series fixes the maximum possible (\emph{cut-off}) frequency to
$\frac{2\pi}{T}L$.  Note that it explicitly omits the zero-frequency term,
which is a desired restriction, in order to fulfill:
$\int_0^T\!\!{\rm d}t\;f(t) = 0\,$.

The manifold spanned by the $f_i$ coefficients is not yet, however, our
parameter space, since we still want to enforce the conditions
$f(0)=f(T)=0$, and fix the fluence:
$\overline{I} = \int\!\!{\rm d}t\; \varepsilon^2(t) = \overline{I}_0\,$.
As discussed in Ref.~\cite{Krieger2011}, these conditions reduce the degrees
of freedom from $2L$ to $2L-2$: the final parameters $u_1,\dots u_{2L-2}$ are
finally the hyperspherical angles that characterize a sphere of constant
fluence, determining the Fourier coeffiencients: $f_i = f_i[u]$.

In all the OCT calculations to be shown below we have fixed the wavelength of the
fundamental frequency $\omega_0$ to 800~nm, a very common value used in
laboratories equipped with a Ti:sapphire source. The total pulse duration is
fixed to 200 cycles, $T=200\;2\pi/\omega_0$, which corresponds to 533~fs
approximately. The envelope function $f(t)$ is then restricted to have
frequencies no larger than $\omega_0/60$. The fluence [Eq.~(\ref{eq:fluence})]
is then fixed to a value (around 5.0~a.u.) that ensures a sufficiently
non-linear response of both the Hydrogen and Helium atoms, while not causing a
substantial ionization. Fixing the fluence does not imply fixing the peak
intensity; however the simultaneous existence of a maximum frequency puts a
limit on it; in practice, the peak intensities observed in the optimal pulses
are in the range of 5 10$^{13}$ - 10$^{14}$ W/cm$^2$.

The optimization are started from randomly generated sets of parameters
$u$. Since the procedure finds local maxima, we have performed several
searches for each case, choosing afterwards the best among them. In order to
have some ``reference'' to compare the optimal run to, we define a reference
pulse as:

\begin{equation}
\varepsilon_{\rm ref}(t) = \varepsilon_0 \cos\left( \frac{\pi}{2} \frac{2t-T}{T} \right) \cos(\omega t)\,,
\label{reference-pulse}
\end{equation}
i.e. a cosinoidal envelope that peaks at $t=T/2$ with a value of
$\varepsilon_0$, chosen to fulfill the fluence condition.

\section{Results}

\begin{figure}
\centerline{
\includegraphics[width=0.7\columnwidth]{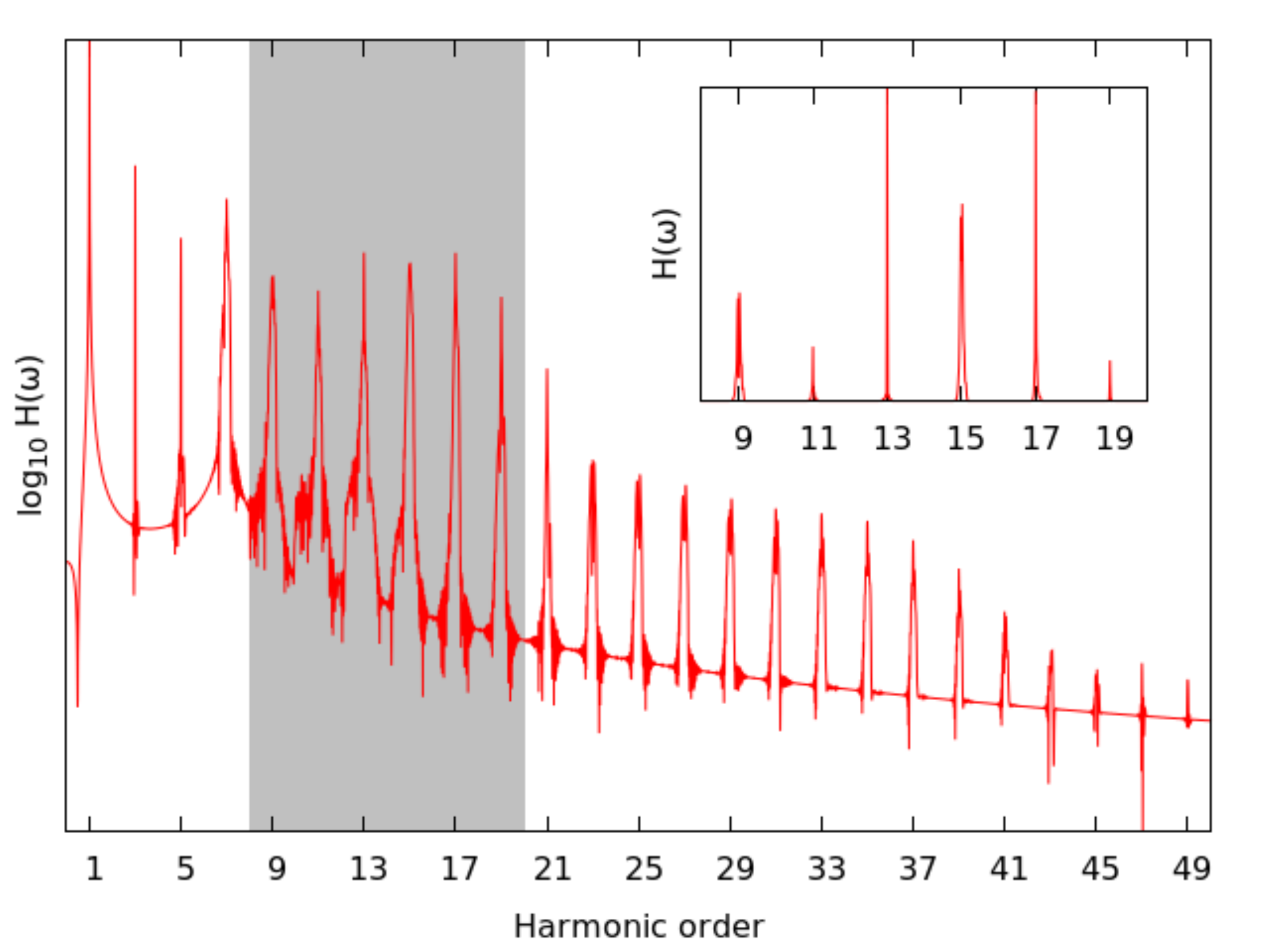}}
\centerline{
\includegraphics[width=0.7\columnwidth]{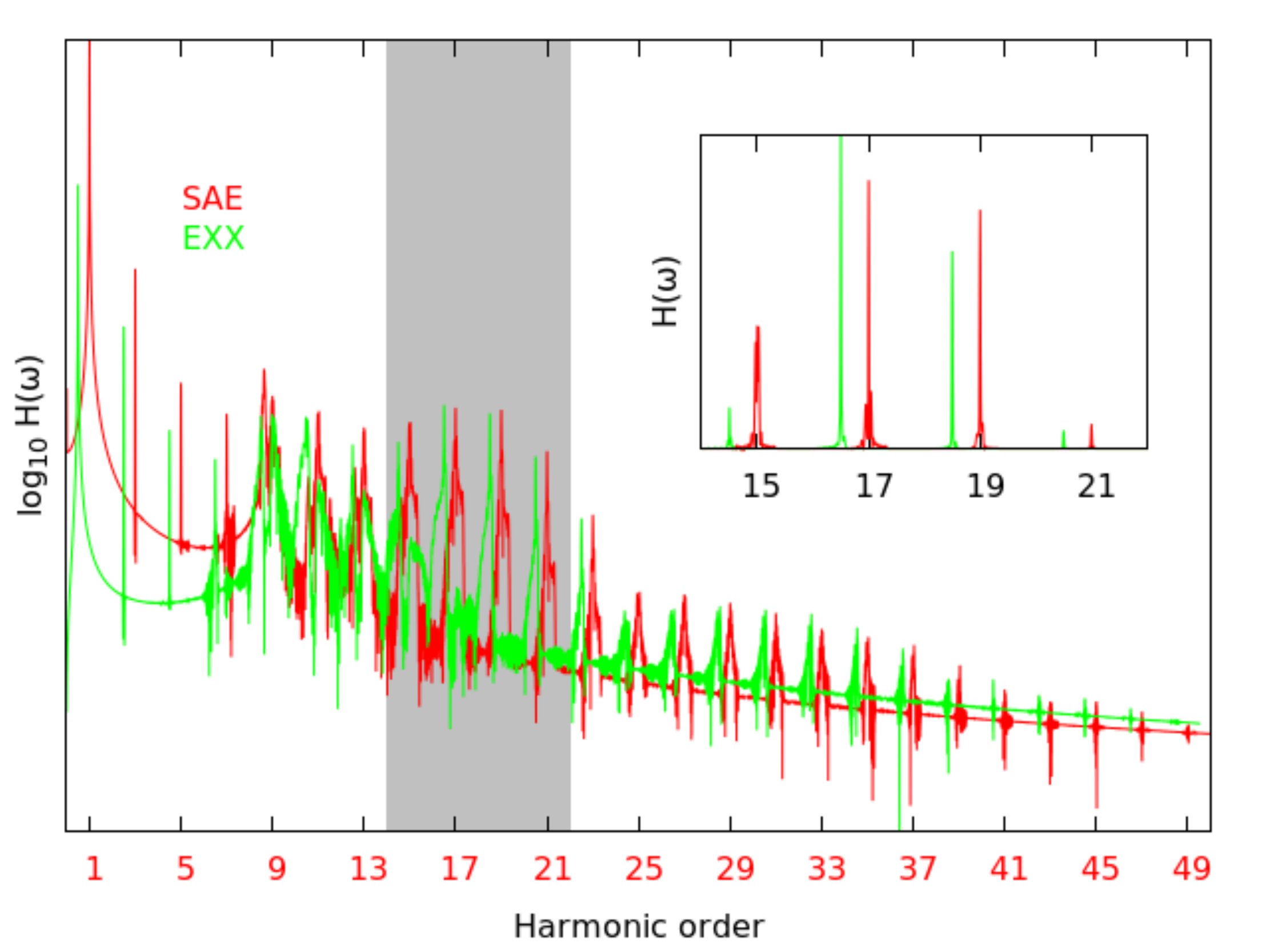}}
\caption{
\label{fig:H-He}
HHG spectrum of the Hydrogen (top) and He (bottom) atoms, with the reference
pulse of Eqn.~(\ref{reference-pulse}.  For the case of the He HHG spectra we
show two results: one solving the TDDFT equations using the EXX functional
(green) and the other solving the single-active-electron (SAE) (red) equation,
commonly used by the strong-field community. To make more clear the comparison
between EXX and SAE we shifted the SAE spectra by 0.5$\omega_0$ in the x-axis. The
shaded area contains the harmonics of interest. This area is also displayed in
the inset, with a linear $y$-axis scale.  }
\end{figure}

The calculated HHG spectrum emitted by the Hydrogen and Helium atoms,
irradiated by the reference pulse, is depicted in Fig.~\ref{fig:H-He}. Note
that there is a range of harmonics with comparable intensities forming a
plateau (9 to 19 in H and 15 to 21 in He).  Because of this, we have selected
that range (shaded in the plot) to perform the selective optimisations. The
range is displayed, this time with a linear $y$ axis scale, in the inset. For
the case of He we show the EXX and SAE results.  As in He two electrons
populate the only orbital in a spin-compensated configuration, the SAE
approximation, in this case, consists in neglecting the interaction between
the electrons during the action of the field, freezing the potential to its
ground state shape. In this adiabatic-DFT context, it amounts to ignoring the
time-evolution of the Hartree, exchange and correlation potentials, and is
useful to gauge the relevance that correlations may have on the HHG
optimisation.

\begin{figure}
\centerline{\includegraphics[scale=0.25]{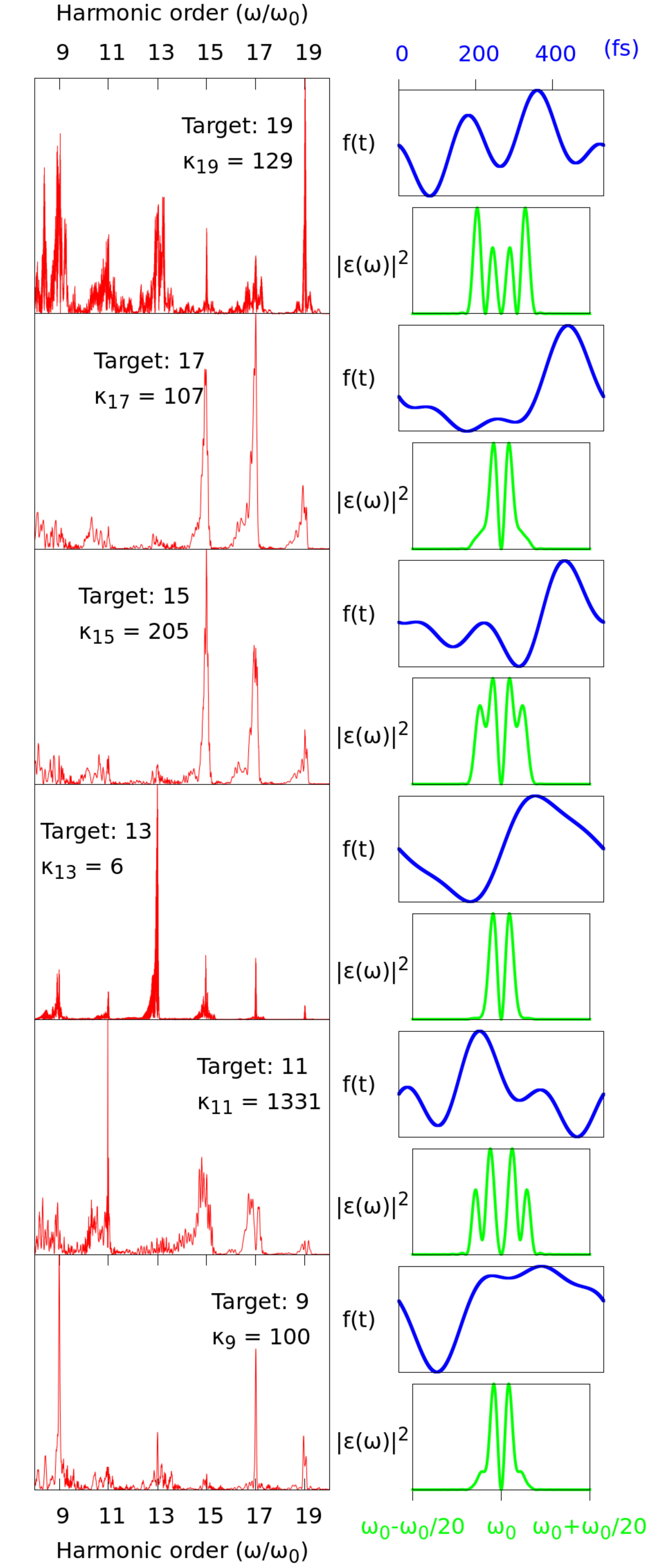}}
\caption{
\label{fig:Hplot}
Optimized HHG spectra (left panels), and corresponding optimal
fields (right panels), for the Hydrogen atom case. The optimal fields are
plotted in the time domain (only the envelope function $f(t)$ is shown),
and in the frequency domain. The HHG spectra are shown in a linear
scale, normalized in each case up the value of the maximun value. The 
enhancement factor defined in Eq.~(\ref{eq:enhancement-factor}) is also shown.
}
\end{figure}

Let's discuss first the case of optimising the HHG spectra of H. We used the
target given by Eq.~(\ref{eq:hhgtarget2}) to optimise the odd orders from the
9th to 19th. To enhance the 9th harmonic, for example, we set $\alpha_9=5$,
and $\alpha_{11}=\alpha_{13}=\alpha_{15}=\alpha_{17}=\alpha_{19}=-1$ (all
other $\alpha_k$ are zero). In this manner, the sum of all coefficients is
zero, avoiding any improvement of the merit function due to a mere overall
reduction or increase of the spectrum. The results are displayed in
Fig.~\ref{fig:Hplot}. From top to botton, in the left panels, the spectra
produced by the optimal fields for the 19th, 17th, \dots, 9th harmonic. In the
right panels, the optimal fields themselves; their envelopes in real
time, as well as their power spectrum.

The resulting fields produce considerably higher harmonic outputs than the
unshaped, reference field. To quantify this point we introduced 
an \emph{enhancement factor} that is displayed in each plot, defined as:
\begin{equation}
\label{eq:enhancement-factor}
\kappa_j = 
\frac{
\max_{\omega \in [k\omega_0 - \beta, k\omega_0
+ \beta]} \lbrace H[\varphi](\omega)\rbrace
}
{
H_{\rm ref}(j\omega_0)
}\,,
\end{equation}
where $H_{\rm ref}$ is the spectrum obtained with the reference field, and $H$
the one obtained with the optimal field (the computation of the max function
is not needed for the former, because due to the regularity of its envelope
function, $H_{\rm ref}$ always peaks at the precise integer multiples
$j\omega_0$).  This enhancement factor greatly vary from case to case (i.e. it
is 6 for $j=13$, and 208 for $j=15$). Note that the plots do not share the
same $y$-scale; they are scaled in each case to the value of the maximum of
the plot.

\begin{figure}
\centerline{\includegraphics[scale=0.25]{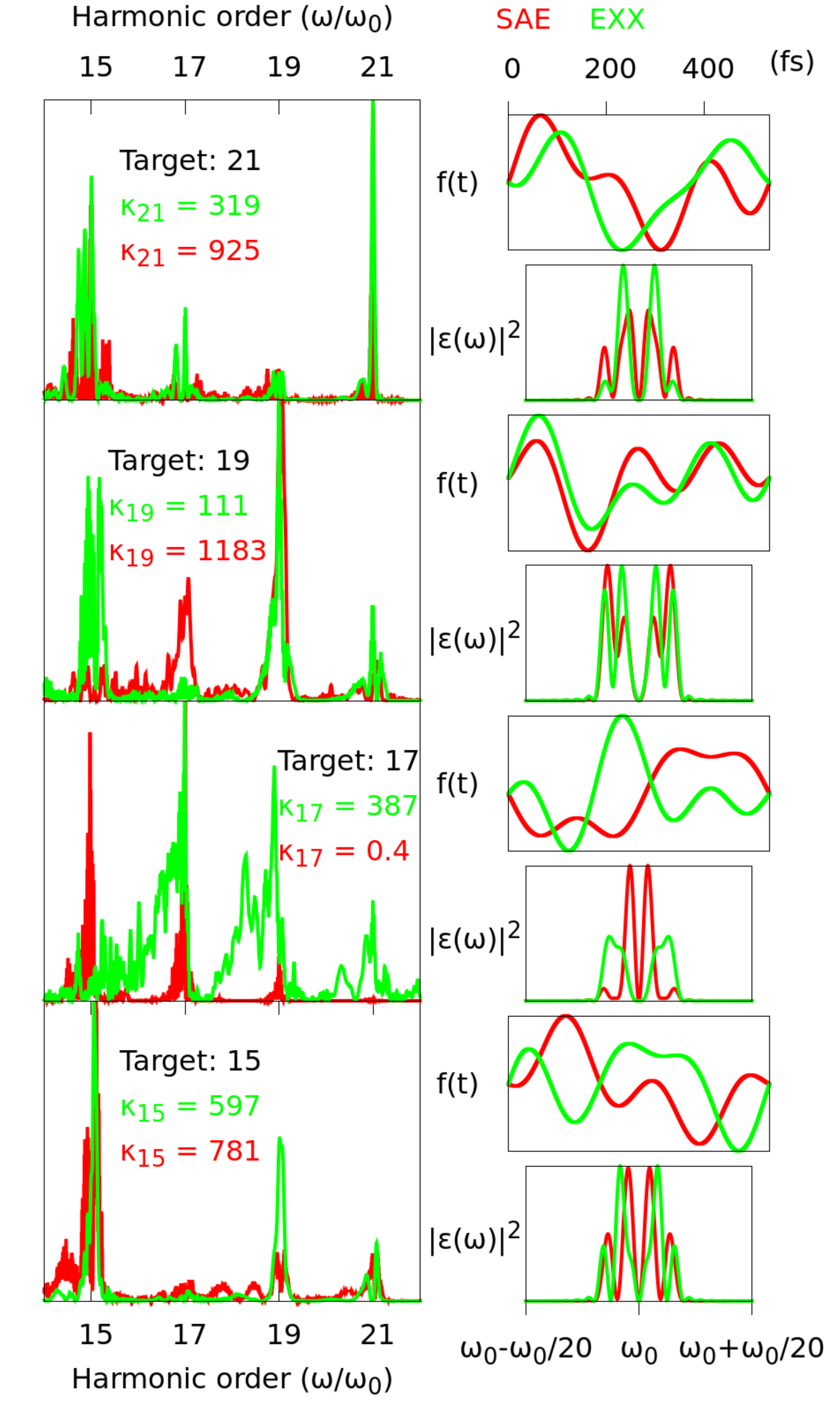}}
\caption{
\label{fig:Heplot}
Optimized HHG spectra (left panels), and corresponding optimal fields (right
panels), for the Helium atom case. As in Fig.~(\ref{fig:H-He}) we show in
green the results within TDDFT using the EXX functional and in red the ones
using the SAE approximation. The optimal fields are plotted in the time domain
(only the envelope function $f(t)$ is shown), and in the frequency domain. The
HHG spectra are shown in a linear scale, normalized in each case up the value
of the maximun value.  The enhancement factor defined in
Eq.~(\ref{eq:enhancement-factor}) is also shown.  }
\end{figure}

We turn now our attention to the case of the Helium atom, that contains two
electrons. The interaction between these is treated here with TDDFT, within
the EXX approximation.  As in the previous case we performed optimisations
based on the target given by Eq.~(\ref{eq:hhgtarget2}), now for the orders
15th to 21st, fixing the coefficients $\alpha_k$ in an analogous manner.  The
results are displayed in Fig.~\ref{fig:Heplot}. From top to bottom, in the
left panels, the spectra produced by the optimal fields for the 21st, 19th,
17th, and 15th harmonic. In the right panels, the optimal fields themselves.

The enhancement factors achieved are quite large, and as in the case of
Hydrogen, rather different from case to case. This rather large enhancement of
the wanted harmonic is not accompanied by a full depletion of the neighbouring
ones -- in fact, they are also increased. This partial selectivity is also
similar to the Hydrogen results. To quantify the role of electron-electron
interactions we show in the same Fig.~\ref{fig:Heplot} the SAE results (red
curve). Qualitatively, the SAE results are not very different to the ones
obtained with the EXX functional, in terms of intensity enhancements and
selectivity.  The fact that the calculated optimal fields and the spectra are
different for both EXX and SAE illustrate not only the intrinsic non-linearity
of the optimisation algorithms and the rather large number of possible local
maxima, but also the fact that electron interaction does play a role in the
generation and optimisation of harmonics.  Indeed, by looking in more detail
to the results shown in Fig.~\ref{fig:Heplot} for the 15th and 19th harmonic
optimisation, we see that EXX with respect to SAE provides a better
selectivity and harmonic enhancement, measured by the height of the desired
harmonic and the quenched of the neighbouring ones. Therefore electron
correlation seems to play a role in the optimisation of harmonics. This fact,
together with the common knowledge that heavier noble gases emit stronger HHG
radiation that light ones (whereas the SAE that predicts similar
spectra)~\cite{PhysRevA.48.4709,PhysRevLett.82.1668,PhysRevLett.96.223902}
support our findings about the limitations of the SAE approximation and the
role of electron interactions. In fact we can expect larger enhancement
factors to be reached by applying the present optimisation techniques to
heavier atomic/molecular systems.

\section{Conclusion}

In conclusion, we have investigated, by theoretical means, the
possibility of tuning the shape of the HHG spectrum of the Hydrogen
and Helium atoms by shaping the slowly varying envelope of a 800 nm,
200-cycles long laser pulses. For this purpose, we have optimised a
functional designed to enhance selected harmonics. The allowed
modifications of the pulse are very constrained, since we enforce a
maximum envelope frequency no larger than 1/60 of the fundamental
frequency. This means very slowly varying envelopes. However, the
picture that emerges of our analysis is that these relatively small
modifications produce strong variations of the spectra, allowing for
significative increases of the harmonic intensities.  These
enhancements are not fully selective, since the neighbouring harmonics
also increase, but to a lesser extent.  The outcome depends of the precise
definition of the target functional, which is a topic to be
investigated further. There is ample freedom to choose this object,
and a different option may yield better selectivity -- while perhaps
reducing the total enhancement, or vice-versa.

The spectra have been computed with a fully quantum mechanical description, by
explicitly computing the time-dependent dipole moment of the systems.  The
results presented here correspond to the single-atom response -- we have not
propagated Maxwell's equations in a atomic gaseous medium. Therefore, this
work demonstrates the relevance of the single atom response for HHG and how
this single-atom response is significantly altered by the envelope of the
laser pulse, even for the small modifications allowed in our scheme. We have
shown that few orders of magnitude HHG enhancement factor can be reached at
the single-atom level. Thus, if this fact is combined with the phase matching
method used for HHG generation we would be able to reach much higher global
harmonic enhancement in atomic and molecular gases. Moreover, our results
illustrate the role of electron-electron interactions in this optimisation and
control of HHG. This can be qualitatively rationalised in terms of the larger
Hilbert space spanned by the interacting system as compare to the simplest
single-active electron scheme (or any other non-interacting electron
approach).

\section*{Appendix: Methods}
\label{Methods}

The propagation of the TDDFT  equations are those of the single-particle
orbitals forming the Slater determinant, a set of equations usually called
``time-dependent Kohn-Sham'' (TDKS) equations:
\begin{eqnarray}
{\rm i}\frac{\partial}{\partial t}\varphi_i(\vec{r},t) & = & -\frac{1}{2}\nabla^2 \varphi_i(\vec{r},t)
+ v_{\rm KS}[n](\vec{r},t)\varphi_i(\vec{r},t)\,,
\\
\label{eq:initial-value}
\varphi_i(\vec{r},0) & = & \varphi_i^{\rm gs}(\vec{r})\,.
\end{eqnarray}
The initial values specifed by Eqs.~(\ref{eq:initial-value}) are given by the
ground-state Kohn-Sham orbitals, computed with static DFT.
The time-dependent density of the system may be retrieved from the KS orbitals with the
simple formula:
\begin{equation}
n(\vec{r},t) = \sum_{i=1}^{N/2}\mu_i\vert \varphi_i(\vec{r},t)\vert^2\,.
\end{equation}
where $\mu_i$ is the occupation of each orbital, which is equal to two if we
consider a spin-compensated system of $N$ electrons, doubly occupying a set of
$N/2$ spatial orbitals $\varphi_i$ ($i=1,\dots,N/2$). 

The potential that appears in those equations, $v_{\rm KS}$ (the ``Kohn-Sham
potential'') is a functional of this density, and is defined as:
\begin{equation}
v_{\rm KS}[n](\vec{r},t) = v(\vec{r}) 
+ \varepsilon(t)\vec{\pi}\cdot\vec{r} + v_{\rm H}[n](\vec{r},t) + v_{\rm xc}[n](\vec{r},t)\,,
\end{equation}
where the Hartree potential $v_{\rm H}$ is given by:
\begin{equation}
v_{\rm H}[n](\vec{r},t) = \int\!\!{\rm d}^3r'\;\frac{n(\vec{r}',t)}{\vert \vec{r}'-\vec{r}\vert}\,,
\end{equation}
and $v(\vec{r})$ is the static external potential.
The time-dependent external potential for these one-electron equations is
given by $\varepsilon(t)\vec{\pi}\cdot\vec{r}$, in terms of objects already defined.

We have studied the two simplest atoms, Hydrogen and Helium. For the Helium
atom, we have used TDDFT with the exact-exchange functional.
 In order to assess the possible relevance of the electron-electron
interaction, we have repeated the Helium atom calculations employing the
single active electron (SAE) approximation, which in this case amounts to
freezing the Hartree, exchange and correlation functional to its ground-state
value during the propagations. In this manner, we are
effectively ignoring the electron-electron interaction during the propagation,
and may gauge the relevance that it may have on the possibility of changing
HHG spectra via smooth variations of the envelope function.

For the purpose of studying the HHG of atoms in linearly polarized pulses,
one-dimensional (1D) models have been routinely employed in the past, and we
have adhered to this practice, since it provides a good qualitative picture,
while substantially reduces the computational cost. The nucleus-electron
interaction has the soft-Coulomb form:
\begin{equation}
v(x) = -\frac{Z}{\sqrt{a^2+(x-x_0)^2}}\,.
\end{equation}
for an electron placed at $x$ and a nucleus of charge $Z$ placed at $x_0$. The
constant $a$ may be tuned to reproduce some atomic property (e.g. ionization
potential), although in this case we have simply fixed it to one for both
Hydrogen and Helium.

Everything has been implemented in the {\tt octopus} code~\cite{marques2003,
  castro2006}. The wavefunctions, potential, densities, etc. are represented
in this code by the values they take at points of a real space grid. The
Laplacian operator, needed to compute the kinetic part of the Hamiltonian, is
computed using a 9-point finite difference formula. The propagations are
performed by dividing the full time interval into short time steps $[t_0, t_1
= t_0+\Delta t, t_2 = t_0+2\Delta t,\dots, T]$, and approximating the
short-time evolution operator $\hat{U}(t_{i+1},t_i)$ with the exponential
mid-point rule:
\begin{equation}
\hat{U}(t_{i+1},t_i) \approx \exp\lbrace -i\Delta t
\hat{H}(t_i+\frac{1}{2}\Delta t) \rbrace\,.
\end{equation}
The action of the exponential on a state vector is computed by making use
of the Lanczos polynomial expansion (see Ref.~\cite{Castro2004} for a
discussion of the propagation schemes used in {\tt octopus}).
The full details about the combination of TDDFT and
QOCT were explained in Refs.~\cite{Castro2012a} and \cite{castro2012}.

\section*{Acknowledgments}
This work was supported by the European Commission within the FP7
CRONOS project (ID 280879). 
AR acknowledges financial support from  the  European Research Council Advanced Grant
DYNamo (ERC-2010-AdG-267374), Spanish Grant (FIS2010-21282-C02-01),
and Grupos Consolidados UPV/EHU del Gobierno Vasco (IT578-13).


\end{document}